\documentclass[twocolumn,showpacs,preprintnumbers,amsmath,amssymb,floatperchapter]{revtex4}
\usepackage{dcolumn}% Align table columns on decimal point
\usepackage{bm}% bold math
\usepackage{epsfig}%
\newcommand{\tb}{\textcolor{black}}
\newcommand{\tr}{\textcolor{black}}
\usepackage[colorlinks]{hyperref}
\usepackage{ulem}
\usepackage{amsmath}
\begin{document}

%\preprint{APS/123-QED}

\title{Experimental study of the magnetic field distribution and shape of domains near the surface of a type-I superconductor in the intermediate state.}% Force line breaks with \\

\author{V. Kozhevnikov$^{1}$,  A. Suter$^{2}$,  T. Prokscha$^{2}$ and C. Van Haesendonck$^3$}
%\vspace{4 mm}
\affiliation{
$^1$Tulsa Community College, Tulsa, Oklahoma 74119, USA\\
$^2$Paul Scherrer Institut, 5232 Villigen PSI, Switzerland\\
$^3$Solid State Physics and Magnetism Section, KU Leuven, BE-3001 Leuven, Belgium
}.\\

%\date{\today}% It is always \today, today,
             %  but any date may be explicitly specified

\begin{abstract}
\noindent 
The importance of accounting for the inhomogeneity of the  magnetic field distribution and roundness of domain walls near the surface of type-I superconductors in the intermediate state for forming  the equilibrium flux structure was predicted by Landau eight decades ago. Further studies confirmed this prediction and extended it to all equilibrium properties of this state. Here we report on direct depth-resolved measurements of the field distribution and shape of domains near the surface of high-purity  type-I (indium) films in a perpendicular field using Low-Energy Muon Spin Rotation spectroscopy. We find that at low applied fields (in about half of the field range of the intermediate state) the field distribution and domains' shape agrees with that proposed by Tinkham. However, for high fields our data suggest that reality differs from theoretical expectations. In particular, the width of the superconducting laminae can expand near the surface leading to formation of a maximum in  the static magnetic field in the current-free space outside the sample. A possible interpretation of these experimental results is discussed. %We speculate that the apparent contradiction of our observations with classical electrodynamics is due to the inapplicability of the standard boundary conditions %Laplace equation in the vicinity of an ``active" superconductor.       

\end{abstract}

%\pacs{74.20.-z, 74.25.Ha, 78.70.Nx}% PACS, the Physics and Astro\Deltanomy
                             % Classification Scheme.,

\maketitle 

\section{INTRODUCTION}

The intermediate state (IS)  in type-I superconductors is a  classical example of a thermodynamically equilibrium systems with spatially modulated phases, where a continuous medium is split into domains of different phases. Such systems with strikingly similar domain patterns are known in a broad variety of physical-chemical formations, in which the pattern constitutes due to competition between various energy contributions in the system free energy \cite{Seul}. The relative simplicity in tuning the domain separation (period of the domain structure) in the IS by varying the applied magnetic field and/or temperature, makes the IS a unique and very interesting object for studies of such systems. %An example of equilibrium in-plane domain structure in a type-I superconductor  are shown in Fig.\,1 \cite{Faber}. 
Recently, being attracted by the beauty of domain patterns (see, e.g., \cite{Faber,Livingston,Huebener}) and long-standing challenges of the IS physics \cite{Shoenberg, Tinkham, Landafshitz_II}, some of us revisited the problem of the IS experimentally. This resulted in the development of a new theoretical model, which consistently addresses properties of the IS in samples of a planar geometry  \cite{IS, MM}. 

Simultaneously, this study made clear the important role of  surface related properties for forming equilibrium characteristics of the IS. Specifically, the role of the out-of-plane field distribution and domain shape (FDDS) near the surfaces through which the flux enters and leaves the sample. Competition  between  the energy contributions arising from these properties (favoring a fine domain structure) on one side and those arising from superconducting (S)-normal (N) interphase boundaries in the sample bulk (favoring a coarse structure) on the other, optimizes and stabilizes the domain structure. In its turn, the latter dictates all other equilibrium magnetic properties. This includes volume fractions of the N and S phases, the shape of magnetization curve, the induction $B$ in the N domains, the critical field of the IS/NS (Normal State) transition $H_{ci}$, \textit{etc.} Being addressed theoretically, these surface related properties of the IS have never been studied experimentally. We report on direct depth-resolved measurements of the FDDS near the surface of high-purity type-I (indium) films in a perpendicular magnetic field using Low-Energy muon Spin Rotation (LE-$\mu$SR) spectroscopy. \tb{This study is a part of a broader project devoted to investigation of equilibrium properties of superconductors; previous results of this project  are published in \cite{MM,IS,Filaments,MixedS} and summarized in \cite{VK}.}

The near-surface properties of the IS were for the first time  considered by Landau in 1937 \cite{Landau_37}. A cross sectional view of Landau's field/domain configuration for an infinite slab in perpendicular field is shown in Fig.\,1A. Assuming that the boundary of a cross section of the S lamina is the line of induction $B$ all the way including the S/N and the S/V (V stands for vacuum) interfaces with magnitude equal to the thermodynamic critical field $H_c$ at the S/N boundary, Landau calculated the shape of rounded corners of the S laminae near the sample  surface. Interestingly, to meet this condition, Landau splits a central field line (the line coming to point $o$ in Fig.\,1A) into two lines ($ocd$ and $oba$), hence challenging the law of magnetic flux conservation %stemming from the absence of magnetic monopoles 
\cite{Tamm,VK}. Soon thereafter Landau admitted the criticism of Peierls and abandoned this model in favor of a so called branching model \cite{Landau_38,Landau_43} (see also \cite{Shoenberg, De Gennes}), in which the N laminae near the surface split into many thin branches so that the flux emerges from the sample uniformly over the whole surface. This branching model was disproved by %in an experimental masterwork of 
Meshkovskii and Shalnikov in 1947 \cite{Shalnikov,Shoenberg}. 
 
Ten years later Sharvin \cite{Sharvin-Sn} for the first time  observed a regular laminar domain pattern in a slab subjected to a tilted field (see also, e.g., \cite{Huebener}). Since this pattern resembled that expected in the  original (non-branching) Landau model, Sharvin used the latter for the interpretation of his results. Ever since, in spite of criticism of Sharvin's interpretation by Faber \cite{Faber}, the results of Landau's calculations of 1937 \cite{Landau_37} are considered as an accurate representation of the FDDS near the surface of samples in the IS \cite{Tinkham,De Gennes,Abrikosov}. 

There are two simplified modifications of Landau's version of  FDDS proposed by Tinkham and Abrikosov. 

Tinkham \cite{Tinkham} assumed that the dominant contribution  in the surface related properties comes from field inhomogeneities extending over a "healing length" $L_h$ outside the sample \cite{heallenght}.  $L_h=(D_n^{-1}+D_s^{-1})^{-1}$, where $D_n$ and $D_s$ are the widths of the normal and superconducting laminae, respectively. (For the case of tube-like domain shape, observed near the IS/NS critical field \cite{Faber},  $D_s$ and $D_n$ correspond to average tube diameter and the distance between, respectively \cite{Tinkham1971}.) Correspondingly, Tinkham neglects the roundness of the laminae corners (\textit{b} and \textit{c} in Fig.\,1A). Tinkham's configuration of the FDDS is shown in Fig.\,1B. This configuration is consistent with images of the IS flux structure observed in \cite{IS} %(see, e.g. \cite{IS,Huebener}) 
and therefore it is adopted in the aforementioned model \cite{IS,MM}. 

It turned out that Tinkham's version of FDDS works  surprisingly well, although it apparently violates basics of magnetostatics  by allowing the existence of field-free regions within the healing spatial layer. Important to note that  $L_h$ in the Tinkham scenario \textit{is not small}. For instance, $L_h$ estimated from the lattice parameter $D = D_n + D_s( \approx 40 \mu$m) measured in 2.5-$\mu$m thick In film at temperature 1.7 K \cite{IS}, is on the order of 1 $\mu$m and, depending on the applied field, can reach up to 10 $\mu$m. %i.e., four times the sample thickness. 
In this regard, in Landau's scenario the field fills the entire outside space, as it is supposed to be the case in magnetostatics, since a static magnetic field (as well as a static  electric field) should not have voids in free space. The latter follows from a theorem of potential, according to which in free space the static field can be zero only at outskirts of a space it occupies \cite{Landafshitz_II}.

\begin{figure}
	\includegraphics[width=0.9\linewidth]{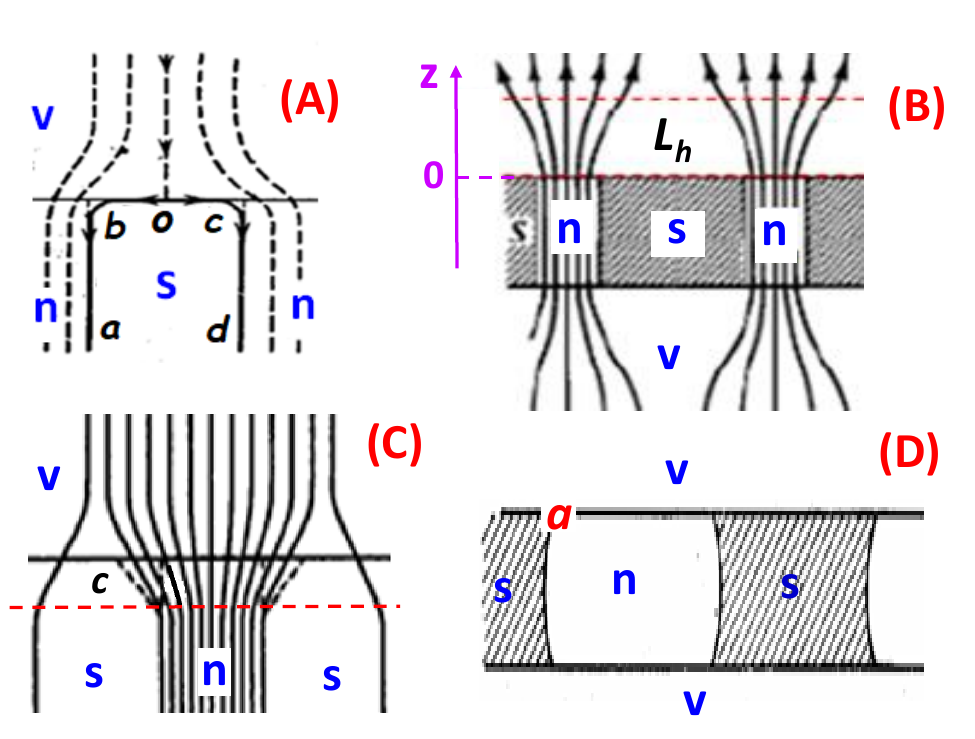}% Here is how to import EPS art
	\caption{Out-of-plane cross-sectional views of the  theoretically predicted field distribution and domain shape near the surface of a planar type-I superconductor in the intermediate state in a perpendicular (A,B,C) and in a tilted (D) magnetic field: (A) Landau \cite{Landau_37}; (B) Tinkham \cite{Tinkham}; (C) Abrikosov \cite{Abrikosov}.  (D) Marchenko \cite{Marchenko}. Blue letters $s$ and $n$ indicate superconducting and normal phases, respectively; $v$ stands for vacuum. In (B) $L_h$, the healing length, is an effective width of a spacial layer with disturbed field; in (C) $c$ is size of the laminae corners. The $z$-axis sketched next to (B) will be used throughout the paper; positive (negative) $z$ means above (underneath) the sample surface. See text for other notations. } 
	\label{fig:epsart}
\end{figure}

Opposite to Tinkham, Abrikosov \cite{Abrikosov} assumed  that the major contribution in the surface related properties is due to the roundness of  laminae corners and therefore neglected the field inhomogeneity outside the sample. However, the latter means that the field near the surface is uniform and therefore this scenario is inconsistent with in-plane images of the IS flux structure. Abrikosov's configuration of the FDDS is shown in Fig.\,1C, where size of the corners $c$ is the same as $L_h$ in Tinkham's scenario.  

An interesting result for possible domain shapes was obtained  by Marchenko \cite{Marchenko}. Like Landau \cite{Landau_37}, Marchenko used conformal mapping to calculate the cross-sectional shape of domains in an infinite slab, but in a tilted field. He found that in a \textit{strongly tilted} field the curvature of the corners can change the sign (compared to that in Figs.\,1A and C) as shown in Fig.\,1D . We note that in this case the field lines should leave the N-domains converging instead of diverging as in Figs.\,1A-C, because bending over a sharp corner (marked $a$ in Fig.\,1D) would take enormous energy \cite{Maxwell}. Therefore, the density of the lines (and therefore the field magnitude)  should pass through maximum somewhere in the free space above the N-lamina. Clearly, such a scenario  does not look possible from the standard viewpoint of the classical electrodynamic.

To conclude this brief overview of theoretical scenarios  for FDDS, we note that none of them is consistent simultaneously with classical electrodynamics and experimental images of the flux structure. To find out the real equilibrium FDDS  near the surface of samples in the IS was the goal of our study, which results are presented below.

\section{EXPERIMENTAL TECHNIQUE AND SAMPLES}

Magnetic properties inside and outside a sample near its  surface can be probed using LE-$\mu$SR spectroscopy, where polarized positive muons $\mu^+$ of tunable energy act as local magnetic microprobes \cite{NL,Sonier,Yaouanc}. Being  embedded inside or stopped outside the sample (coated for this purpose by an overlayer of a suitable material) in a site with an average microscopic field (\textit{i.e.,} the induction) $B$, the muon spin precesses with an angular frequency $\gamma_\mu B$, where $\gamma_\mu$ is the muon  gyromagnetic ratio. The muon is a radioactive particle with a lifetime  $\tau_\mu=$ 2.2 $\mu s$. It decays into a positron and two neutrinos. The former is preferentially  emitted in the direction of the muon's spin at the decay instant (with an asymmetry close to 30\%). On the other hand, the spin of a muon stopped in a site where $B=0$ does not precess and the positron is emitted preferentially in the direction of the initial spin polarization  $\textbf{P}(0)$. 

A time-differential $\mu$SR experiment on a quasi-continuous  beam works the following way: a muon counter starts a clock which is stopped by the corresponding decay positron counter. The resulting time difference is recorded in a histogram. Accumulating $10^6-10^7$ such events results in a muon decay histogram, which can be written as

\begin{equation}\label{eq:Nt}
N(t) = N_0\,e^{-t/\tau_\mu} \left[ 1 + A_0 P(t) \right] + N_{\rm bkg},
\end{equation}
where $N_0$ is the scale of recorded decay positrons, $N_{\rm  bkg}$ is a flat background due to uncorrelated events, and $A_0 P(t)$ is the asymmetry signal called a time spectrum containing the relevant information for the experiment. 

The Fourier transform of the time spectrum contains the  magnetic induction distribution $p(B)$ over the muon sites; the peak in $p(B)$ corresponds to the most probable induction in these sites.  

Typically multiple positron counters are utilized in a  $\mu$SR instrument. The positron detectors arrangement in the LE-$\mu$SR setup used in this study is schematically depicted in Fig.\,2.  

\begin{figure}
	\includegraphics[width=0.5\linewidth]{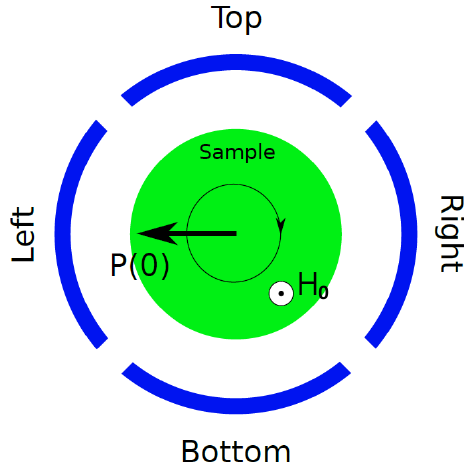}% Here is how to import EPS art
	\caption{Schematic diagram of the LE-$\mu$SR apparatus,  showing the sample (green), the direction of the applied field $\textbf{H}_0$, the initial direction of muon spin $\textbf{P}(0)$ (black arrow), the angular velocity of precessing spin (circular arrow), and the positron counters (blue segments).} 
	\label{fig:epsart}
\end{figure}

The asymmetry signals decay with time due to depolarization  of the muon spin ensemble caused by (a)  microscopic currents and nuclear spins near the muons' sites and (b) a possible gradient of the induction $B$ over the range of the muons' stopping distribution (see \cite{Eckstein,Morenzoni_2002} and Appendix below). The random character of the former leads, within a good approximation, to a Gaussian distribution of the probing $B$; in such case treatment of the $\mu$SR spectra is essentially model-free. Contrarily, in the latter case the field distribution is non-Gaussian  and an adequate theoretical model is required for quantitative interpretation of the spectra  (see, e.g., \cite{NL,Andreas}). However even at very large field gradients (like those within the penetration depth of the extreme type-I materials) the Gaussian approximation yields consistent semi-quantitative results \cite{NL}.

In case of a two-component medium consisting of  domains/regions with zero and non-zero $B$, both precessing and non-precessing asymmetry signals can be recorded at the same time. The initial amplitudes of these signals are proportional to the number of muons stopped in each of these domains/regions and therefore proportional to their volume fraction at a specific depth beneath or height above the surface. Thus, $\mu$SR spectroscopy allows one to measure simultaneously both $B$ in domains/regions where it is non zero and the volume fraction of these regions. Using LE-$\mu$SR, these characteristics can be measured versus distance on both sides of the sample surface by changing the implantation depth (via tuning the muon kinetic energy) inside and the height at which muons are stopped outside the sample.
 
\begin{figure}
	\includegraphics[width=0.9\linewidth]{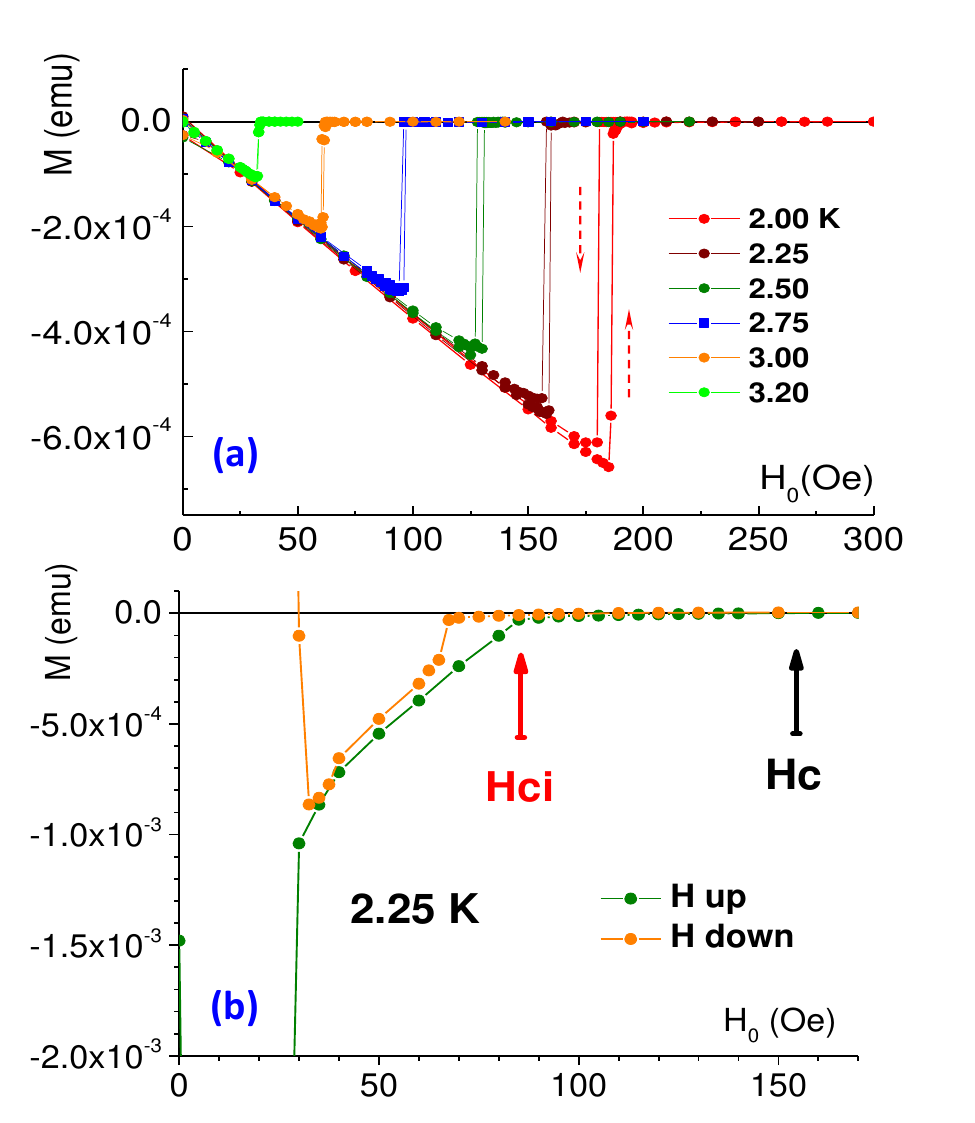}% Here is how to import EPS art
	\caption{Data for the magnetic moment of a sample with  the In-C film. (a) The data measured in the field $\textbf{H}_0$ parallel  to the film at increasing [zero-field cooled (ZFC) sample, arrow up for 2.0 K] and decreasing [field cooled (FC) sample, arrow down] field magnitude  at indicated temperatures. (b) The data obtained in $\textbf{H}_0$ perpendicular to the film at temperature 2.25 K; green points represent the data measured at the increasing  field in ZFC sample, and orange points are data measured at the decreasing field in FC sample; $H_c$ is the thermodynamic critical field determined from the data shown in the upper panel, and $H_{ci}$ is the critical field of the IS/NS transition in the perpendicular field.  } 
	\label{fig:epsart}
\end{figure}

\begin{figure}
	\includegraphics[width=0.87\linewidth]{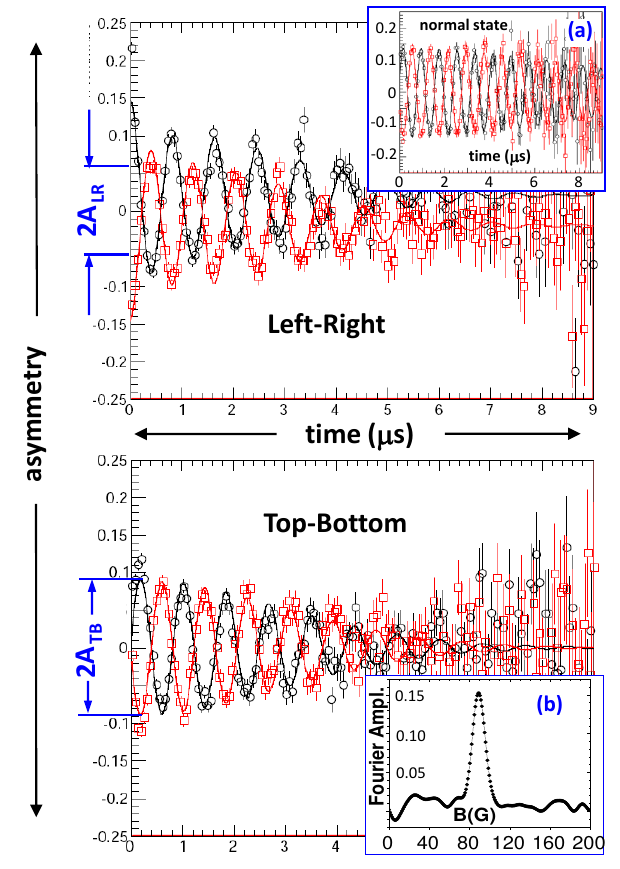}% Here is how to import EPS art
	\caption{Typical time spectra measured outside the  samples. Shown are the spectra of the In-A sample at $T= 2.47$ K and an applied field 68 Oe at a distance $z=$ 330 nm from the sample surface (with 500-nm-thick layer of solid N$_2$). The upper panel shows the spectra recorded on the Left and Right (black and red points, respectively) counters facing to the direction parallel to the  initial polarization $\textbf{P}(0)$. The lower panel shows the spectra recorded on the Top (black points) and Bottom (red points) counters. The black and red lines are fitting curves obtained using Eq.\,(2) (lower panel) and Eq.\,(3) (upper panel); measured $B=89.4(1)$ G. $A_{\rm TB}$ and $A_{\rm LR}$ are obtained from the fit the initial asymmetries due to precessing and non-precessing muons, respectively. Inserts: (a) Time-spectra measured with the sample in the normal state (at $H_0=$ 93 Oe); (b) A phase corrected real part of the Fourier transform  of the time spectrum recorded on the Top-Bottom detectors representing the spectrum of the induction in the regions with non-zero $B$.}
	\label{fig:epsart}
\end{figure}

\begin{figure}
	\includegraphics[width=0.9\linewidth]{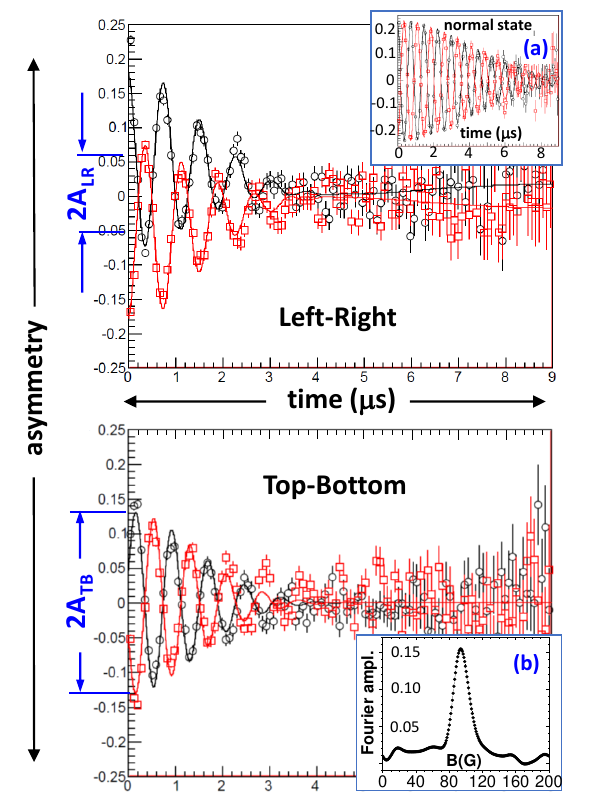}% Here is how to import EPS art
	\caption{Typical time spectra measured inside the  samples. Shown are the spectra taken with In-A sample at temperature $T=2.47$K and an applied field $H_0=63$\,Oe with a muon energy $E=24$\,keV. Corresponding average stopping distance of the muons is 115 nm beneath the surface; measured $B=95.2(2)$\,G. Inserts:  (a) The time spectra measured with muons of the same energy when the sample is in the normal state (at $H_0=98$\,Oe); (b) The phase corrected  real part of the Fourier transform  of the time spectrum shown at the lower panel representing the spectrum of the induction in the normal domains. See captions of Fig.\,4 for more details.}
	\label{fig:epsart}
\end{figure}

If the distribution of the measured $B$ over the stopping range is Gaussian, the asymmetry spectra recorded with the Top and Bottom (TB) and Left and Right (LR) counters (see Fig.\,2) with a sample in the IS have the form   
\begin{eqnarray}
A_0P(t)_{\rm TB} &=& A_{\rm TB}e^{-(\sigma_{\rm TB}t)^2/2} \cos(\gamma_\mu B t+\phi) \label{eq:asymmetry-TB} \\
A_0P(t)_{\rm LR} &=& A_0P(t)_{\rm TB}+ \nonumber \\ \label{eq:asymmetry-LR}
& & A_{\rm LR}[\frac{1}{3}+\frac{2}{3}(1-[\sigma_{\rm LR}t]^2)e^{-(\sigma_{\rm LR}t)^2/2}], 
\end{eqnarray} 
where $A_0P(t)_{\rm TB}$ is the asymmetry recorded vs time  $t$ by the Top and Bottom counters; this asymmetry is caused only by ''precessing muons`` (\textit{i.e.,} muons stopped in domains/regions with non-zero $B$), and $A_{\rm TB}$ is its initial amplitude. $A_0P(t)_{\rm LR}$ is the asymmetry recorded by the Left and Right counters; it is  caused by both precessing and ''non-precessing`` muons, and $A_{\rm LR}$ is the initial amplitude of the asymmetry related to non-precessing muons, \textit{i.e.}, to muons stopped in domains/regions with $B$ = 0; $\sigma_{\rm TB}$ and $\sigma_{\rm LR}$ are rates of depolarization of precessing and non-precessing muons, respectively; and $\phi$ is the initial phase of the muon spin in respect to each counter.

The second term in Eq.\,(3), referred to as Gauss  Kubo-Toyabe function \cite{Kubo,Yaouanc}, describes the depolarization of the non-precessing muons; it  originates from the microscopic field distribution averaging at $B=0$. For the Left, Top, Right and Bottom counters the initial phase $\phi$ equals $0$, $\pi/2$, $\pi$ and $3\pi/2$ respectively. So, the first term in Eq.\,(3) differs from the asymmetry in Eq.\,(2) by the value of $\phi$.

For spectra measured inside the sample normalized amplitudes  of asymmetries 
$A_{\rm LR}/(A_{\rm TB}+A_{\rm LR})$ and $A_{\rm TB}/(A_{\rm  TB}+A_{\rm LR})$  represent volume fractions of the S-component $\rho_s= w_s/w$ and the N-component $\rho_n=w_n/w$, respectively. Here $w_s$ and $w_n$ are, correspondingly, volumes of superconducting and normal phases in a slice  parallel to the film surface and having the thickness equal to the width of the stopping distances distribution of the implanted muons of given energy; and $w\equiv w_s + w_n$ is the volume of the entire slice. When measured outside the sample, $\rho_s$ and $\rho_n$ are the volume fractions of the regions with zero and non-zero induction, respectively.

For samples in the NS the asymmetries recorded on all  counters have the form of Eq.\,(2), \textit{i.e.,} they differ from each other by the initial phase only, since for the used setup the solid angle and the efficiency of all the positron counters is identical.

The field inside the sample at different distances (depths)  from the surface was probed in the standard LE-$\mu$SR  way, \textit{i.e.,} by implanting muons with different energies in the range from 3 to 25 keV. Corresponding average stopping distances for In range from 20 to 140 nm, respectively (see the Appendix).

To stop muons outside the sample we used a layer of nitrogen  deposited on the sample surface from the vapor phase; muons were implanted and stopped in this layer. The rate of N$_2$ deposition is determined by the sample  temperature and pressure of nitrogen gas filling the cryostat. In our case the rate was close to 50 nm/min. Then the thickness of the N$_2$ layer is determined by the deposition time, \textit{i.e.,} by the time during which the cryostat is filled with nitrogen. Upon completing measurements with one layer, it was removed by heating the sample to $\sim$30 K. Afterwards the sample was cooled back to the original temperature and a new nitrogen layer was deposited.  In all these ''outside`` measurements, the energy of the muons was 14.3 keV; the average muon stopping depth in the N$_2$ layer was 170 nm, as calculated with the program TRIM.SP \cite{Eckstein,Morenzoni_2002}. A graph for stopping distances of muons in solid nitrogen is provided in the Appendix below.

Muons stopping in solid nitrogen may capture an electron to  form the hydrogen-like muonium state. The precession frequency of muonium is about hundred times faster compared to the precession frequency of the muon, and cannot be observed in the field range for the current LE-$\mu$SR setup. In the deposited N$_2$ layer the fraction of muons precessing at its Larmor frequency is about 40 to 50\% \cite{Prokscha_2007}, causing a corresponding reduction of the amplitude $A_{\rm TB}$ of the precession signal.

We used two indium film samples In-A and In-C. Each film was  deposited  on a polished sapphire disc of 60 mm in diameter. Simultaneously a few smaller size samples were fabricated for the film characterization.  The thickness (residual resistivity ratio RRR) of the In-A and In-C films is 3.86 $\mu$m (610) and 2.88 $\mu$m (570), respectively. \tb{The thickness of the In-A film was measured using optical interference profiler, and the thickness of the In-C film was measured with calibrated ultrasonic sensor.} The elastic mean path is 12 and 11 $\mu$m for the In-A and In-C films, respectively. The mean free path was calculated from the measured RRR with use of data \cite{Serin} for the product of the mean free path and resistivity at room temperature \cite{purity}. The film In-A was the same film which was used as In-A sample in \cite{MM} (\textit{i.e.,} samples used in \cite{MM} and in this work were deposited simultaneously). Details for the films fabrication and a typical image of the film surface are available in \cite{NL}.

Representative data for the magnetic moment $M$ of the In-C  film measured in parallel and perpendicular fields are shown in Figs.\,3a and 3b, respectively. $M$ was measured  using a  Quantum Design dc magnetometer (Magnetic Properties Measurements System).  Similar data for the In-A film are available in  \cite{MM}. 

Apart from a supercooling region at the S/N transition,  magnetization data for both films were fully reversible when measured in parallel field. We remind that, like in all other first order phase transitions (\textit{e.g.,} melting, boiling, \textit{etc.}),  the supercooling effect is caused by the positive interphase energy and presence of this  effect is a hallmark of the high purity of the substance under study \cite{VK}.

In perpendicular field the magnetization data were reversible  in the fields from $H_{ci}$ down to about 0.3$H_{ci}$. %, where $H_{ci}$ is the critical field of the IS/N transition. 
This is the field range in which most of the LE-$\mu$SR  spectra were taken. Hence, we conclude that the spectra were measured with nearly pinning free samples and therefore  results reported below represent thermodynamically equilibrium properties of the IS. 

\begin{figure}
	\includegraphics[width=0.9\linewidth]{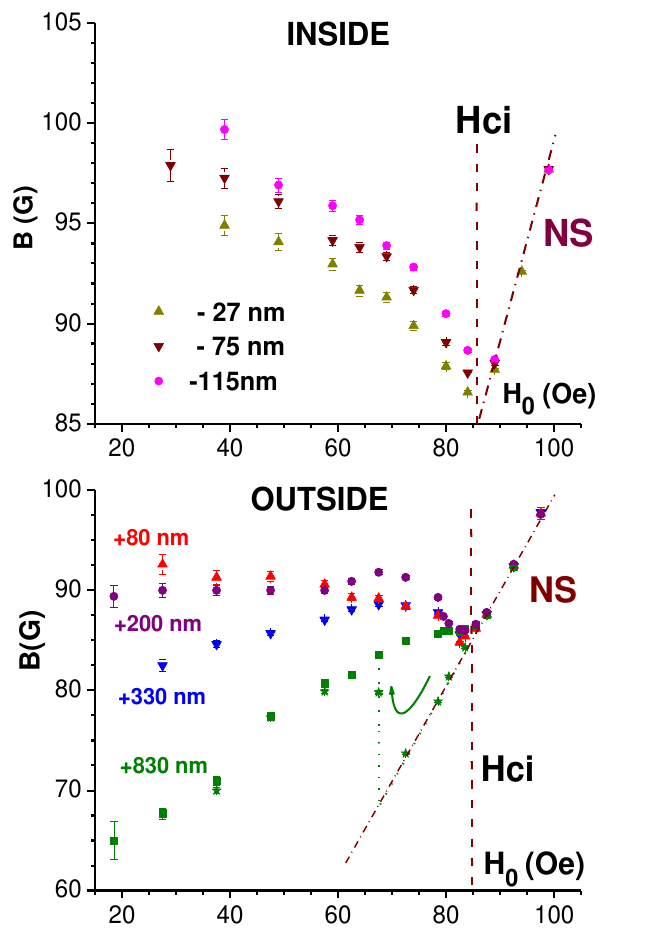}% Here is how to import EPS art
	\caption{The induction $B$ inside (upper panel) and  outside (lower panel) of the In-A sample measured at temperature 2.47 K vs applied field $H_0$ at indicated distances from the surface. Negative distances (in (a)) are the depths beneath the surface; positive distances (in b)) are the heights above the surface. $H_{ci}$ is the critical field of the IS/NS transition; dashed-doted line labeled by NS is the graph $B(H_0)$ for the normal state, where $B=H=H_0$ ($H$ is the field intensity).} 
	\label{fig:epsart}
\end{figure}

Before describing the experimental results we note that a  problem similar to that we discuss here was addressed in \cite{Niedermayer} for the mixed state in an extreme type-II superconductor (YBCO film) in perpendicular field. It was a first application of the LE-$\mu$SR technique to superconductivity, targeted to demonstrate the rich capabilities of the new technique. The experiment was performed at a single temperature (20\,K) and field (104 Oe) by changing the energy of  muons implanted in the film and in a thin silver layer deposited on an identical film in order to stop muons outside the sample. The Tinkham's formula mentioned above was used to interpret the obtained $\mu$SR spectra. It was found that the use of this formula leads to consistency of the measured spectra with calculations based on the London model for the mixed state.

\section{EXPERIMENTAL RESULTS}

The LE-$\mu$SR experiments were performed at the LEM beamline  of the Swiss Muon Source at the Paul Scherrer Institute~\cite{Prokscha}. In all measurements the cryostat was kept at base (the lowest) temperature. The temperature of each sample was  determined \textit{in situ} using the sample's phase diagram  $H_{ci}(T)$ obtained from the magnetization data measured in perpendicular field. It was $T=2.47$K ($T=2.24$K) for the In-A (In-C) film. In all but one run the samples were cooled in zero applied  field (i.e., in the Earth field) and the measurements were conducted at increasing field. The measurements with a 1-$\mu$m-thick N$_2$ layer were performed at increasing and decreasing field as described below. The number of positrons (\textit{i.e.,} the number of implanted muons) collected for each experimental point was 4$\cdot$10$^6$ (to save beamtime some spectra in the N state were taken with number of muons reduced to 2.5$\cdot$10$^6$). All the $\mu$SR data were fitted using \texttt{musrfit}  \cite{A-W}. LE-$\mu$SR experiments with In-A and In-C films were performed at different beam cycles with an interval of two years. 
\begin{figure}
	\includegraphics[width=1.0\linewidth]{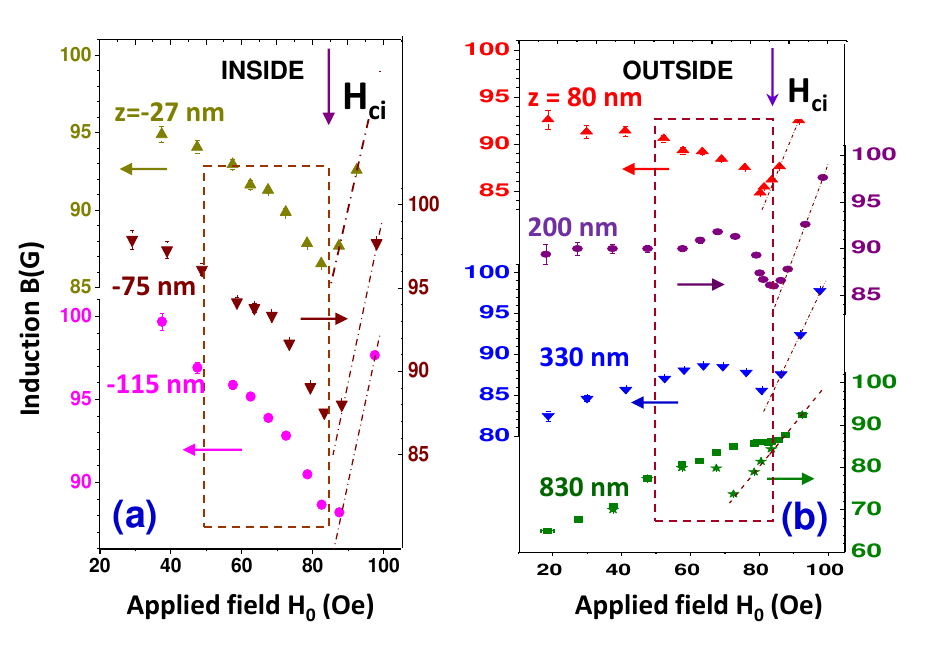}% Here is how to import EPS art
	\caption{The same as in Fig.\,6 data for $B$ vs $H_0$  shown with the shifted vertical scales  as indicated by arrows.}
	\label{fig:epsart}
\end{figure}

Fig.\,4 shows typical time spectra originating from  measurements where the muons were 
 stopped outside the sample (shown are the spectra taken for  the In-A film with a 500 nm thick N$_2$ layer) recorded at Left and Right (upper panel) and Top and Bottom (lower panel) counters in an applied field $H_0=68$ Oe. For comparison, the insert (a) shows the spectra at Left-Right detectors when the sample is in the N state at the same temperature, \textit{i.e.,} at the field ($H_0=98$ Oe) exceeding the critical field of the IS/NS transition $H_{ci}$ ($\approx 85$ Oe). Insert (b) shows Fourier transform of the time spectra for the Top and Bottom counters representing the  $B$-spectrum [$p(B)$] in regions with non-zero $B$. 
 
Typical time spectra measured inside the sample are presented  in Fig,\,5. Shown are the spectra taken for the In-A film with muons accelerated to an energy of 24 keV. Corresponding the average depth at which the $\mu$SR properties are probed is 115 nm. For comparison,  the spectra taken with muons of the same energy when the sample is in normal state are shown in the insert (a). The insert (b) shows the spectrum of induction in the N-domains for ($H_0,E$)=(63 Oe, 24 keV). 

Representative data for depolarization rates  $\sigma_{\rm TB}$ and $\sigma_{\rm LR}$ are shown in Appendix. 

From Figs.\,4 and 5 we see that (i) the initial asymmetry  $A_{\rm LR}\neq 0$ in both cases. This indicates that regions/domains with $B=0$, are present inside (Fig.\,5) as well as  \textit{in the current-free space outside}  the sample (Fig.\,4); and (ii) the $B$-spectra (shown in Figs.\,4b and 5b) are fairly close to the Gaussian field distribution. The latter justifies the use of Eqs.\,(2) and (3) to fit the measured spectra.

Fig.\,6 shows data  for $B$  vs $H_0$ obtained at fixed  distances $z$ beneath (Fig.\,6a, $z<0$) and above (Fig.\,6b, $z>0$) the surface for the In-A film. The muon energies at the "inside" measurements were 5, 16 and 24 keV, corresponding to average depths of 27, 75 and 115 nm, respectively. Measurements at positive $z$ were conducted using four N$_2$ layers with thickness 250, 375, 500 and 1000 nm; distances from the sample surface were 80, 200, 330 and 830 nm, respectively. For clarity, in Fig.\,7 the data shown in Fig.\,6 are presented with shifted vertical axis. 
\begin{figure}
	\includegraphics[width=0.85\linewidth]{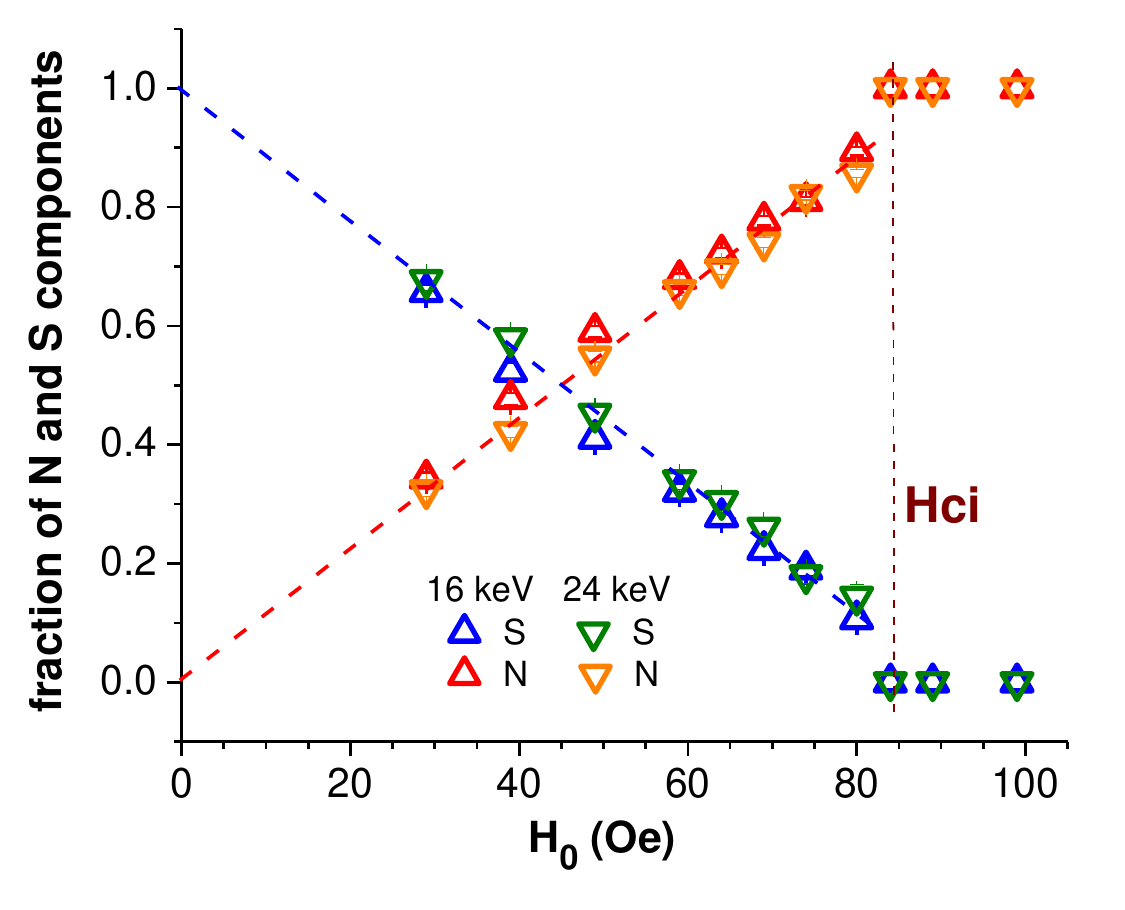}% Here  is how to import EPS art
	\caption{Volume fractions of the superconducting (S) and  the normal (N) components vs applied field at fixed muon implantation energies 16 and 24 keV corresponding to average distances 75 and 115 nm, respectively,  beneath the surface of the In-A film. $H_{ci}$ is the critical field of the IS/NS transition. }
	\label{fig:epsart}
\end{figure}

\begin{figure}
	\includegraphics[width=0.7\linewidth]{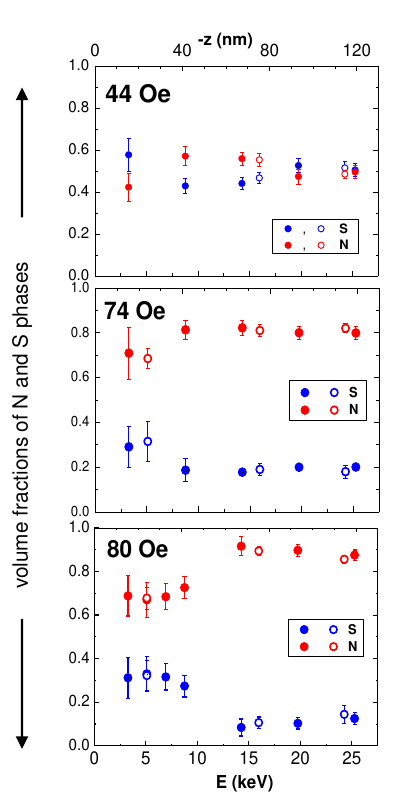}% Here  is how to import EPS art
	\caption{Volume fractions of the superconducting (S) and  the normal (N) components vs muon energy and average distance $z$ in the In-A film at three different values of the applied fields as indicated. The implantation energy of the muons is presented at the bottom scale of each panel; corresponding average distances $z$ are shown at the upper scales. Solid circles are data obtained from spectra measured at fixed fields vs muon energy; open circles are data obtained from spectra measured at fixed energies vs applied field.}
	\label{fig:epsart}
\end{figure}        

The spectra with 1-$\mu$m thick N$_2$ layer ($z=$ +830 nm)  were measured both at increasing field after cooling the sample  in zero applied field and at decreasing field starting from the normal state, \textit{i.e.} with the sample cooled in a field $H>H_c$. The data points for $B$ obtained from these spectra are shown in Figs.\,6b and 7b as green solid squares for increasing $H$ and as green stars for decreasing field. As one can see, (a) the data obtained are well reversible and (b) there is a deep supercooling effect at decreasing field. These results confirm  that the sample was essentially pinning-free in the field range studied. A similar supercooling effect, testifying that IS/NS transition is a phase transition of the first order, was reported in other studies of the IS performed with high purity samples, \textit{e.g.} in  measurements of the electrical resistance \cite{Andrew}, magnetization \cite{MM,Desirant} and $\mu$SR spectra \cite{Egorov}.

As seen from Figs.\,6 and 7,  in the field  range  $0.6\lesssim H/H_{ci} \leqslant 1$ (marked by a dashed rectangle in Figs.\,7a and 7b)  a slight irregularity in $B(H_0)$ inside the sample develops into a strong anomaly outside. The induction $B$ measured with a 250-nm N$_2$ layer ($z$ = 80 nm) monotonically decreases with increasing $H_0$ in a similar way as it takes place inside the sample. This signals that the flux pattern at this distance is about the same as that  near the surface inside. However, $B(H_0)$ measured with a twice as thick N$_2$ layer ($z$ = 330 nm) is non-monotonic: in a major part of the field range the slope of $B(H_0)$ is positive, then it changes sign and above 70 Oe $B(H_0)$ is the same as that for $z=$ 80 nm. One could expect that at the middle between these distances $B(H_0)$ is close to the mean of these two dependencies, \textit{i.e.,} $B(H_0)$ is about  field independent until it meets the first two curves (for $z=$ 80 and 330 nm). We see that $B(H_0)$ at $z=$ 200 nm is indeed close to constant at $H_0\lesssim$ 60 Oe but then it \textit{rises up}  becoming greater than that at both smaller (80 nm) and larger (330 nm) distances. At even larger distance ($z=$ 830 nm) $B(H_0)$ is monotonic again, but now its slope is positive, however $B(H_0)$ is still far from that for the applied field, where $B=H_0$ as shown by the dash-dotted line in Figs.\,6 and 7.        

In all theoretical scenarios for the transverse applied field  (see Figs.\,1A to 1C) the induction $B(z)$ above the N-domains decreases with increasing $z$ starting from $z=0$. However, according to experimental data obtained at the high field ($\gtrsim 0.6H_{ci}$), $B(z)$ first increases before it gradually decreases to the value of the applied field $H_0$ \textit{far away} from the sample (\textit{i.e.,} at $z$ on the order of a few microns). Therefore, the function $B(z)$ outside the sample passes through a maximum implying that the field lines exit the N-domains \textit{converging}, as it can be expected if the cross sectional domain shape is similar to that shown in Fig.\,1D. 

Here it is important to stress that the width of  muon  stopping distribution (see Fig.\,14 in Appendix) is an order of magnitude less than the the scale of the average distances from the sample surface in the "outside" measurements (positive $z$). Therefore the observed variation of the field in the out-of-plane direction cannot be an artifact associated with the finite width of the stopping distribution.  

\begin{figure}
	\includegraphics[width=0.8\linewidth]{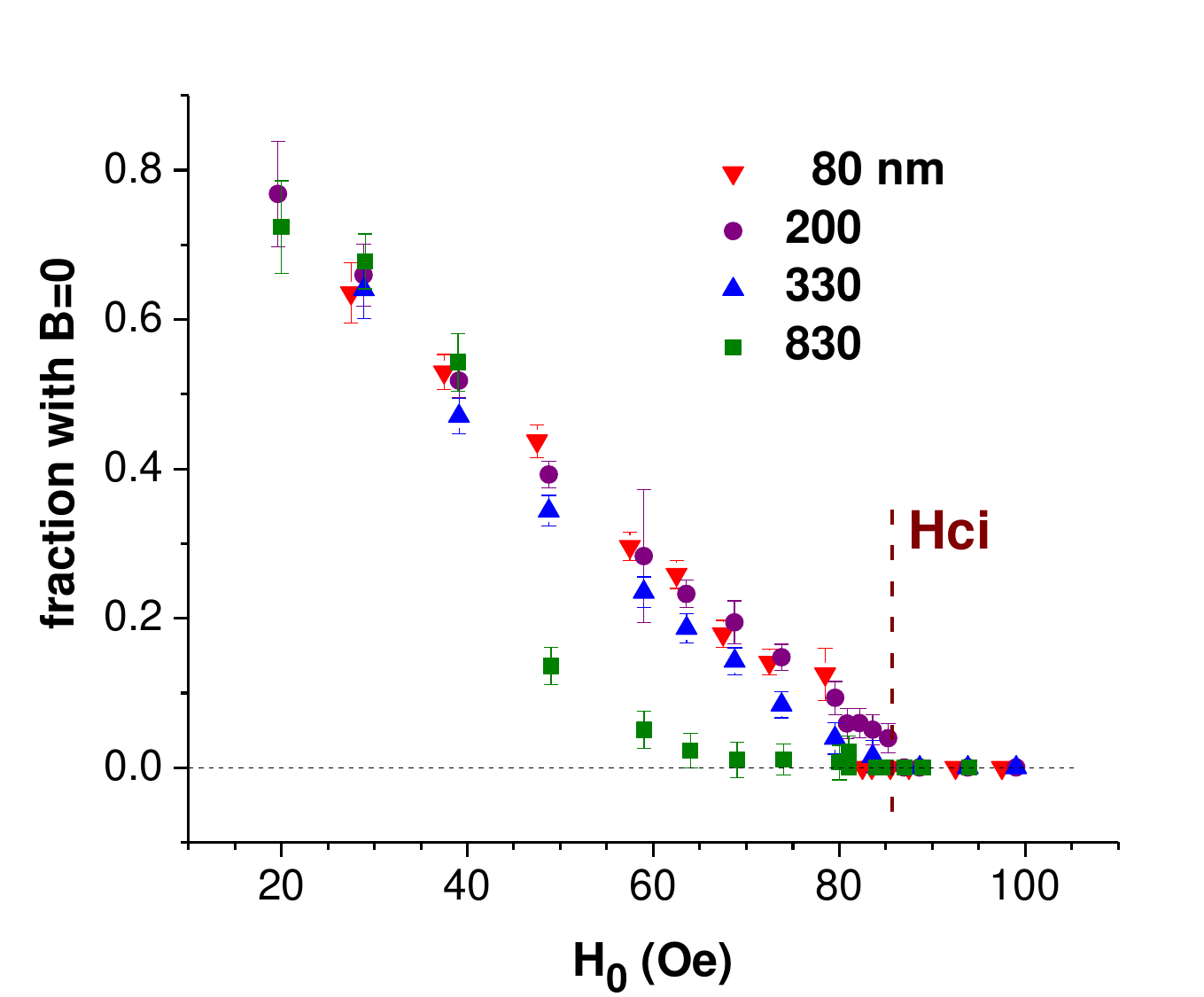}% Here  is how to import EPS art
	\caption{Volume fractions of regions with $B=0$ outside  the In-A film at indicated distances $z$ from the film surface. The dashed line marks $H_{ci}$, the critical field of  the IS/NS transition. All data points were obtained with the sample cooled in zero field.}
	\label{fig:epsart}
\end{figure}

Now we turn to the volume fractions of the  components/regions. Note, that bending of the field lines on both sides of the surface may effect the amplitude of asymmetries and hence the values of $\rho_n$ and $\rho_s$. However, this effect is small \cite{Niedermayer,bending} and does not exceed the error bars for these quantities.

Fig.\,8 shows graphs for $\rho_n$ and $\rho_s$ vs $H_0$  extracted from the spectra measured with muons of 16 and 24 keV corresponding to the average distances $z=$ - 75 and -115 nm, respectively.  We see that the graphs $\rho_n(H_0)$  [$\rho_s(H_0)$]  are close to linear and their extrapolation to $H_0=0$ passes through the origin [unity], hence satisfying the limiting cases $\rho_n=$ 0 [$\rho_s=$ 1] at $H_0=$ 0. A linear dependence of the initial amplitude  of $\mu$SR asymmetry vs applied field measured deeply inside a single crystal tin sample was also reported in  \cite{Egorov}. The linearity of these graphs is consistent with the linear dependence of resistance $R$ vs $H_0$ measured in a spherical sample in direction perpendicular to $\textbf{H}_0$ \cite{Shubnikov} and in cylindrical samples in perpendicular $\textbf{H}_0$  \cite{Andrew}. In a film sample in tilted field $R$ is also a linear function of $H_{0\perp}$ (perpendicular component of $H_0$) \cite{IS}. The resistance of samples in the IS is due to the presence of the N component, directly measured in this work.  Therefore the graphs in Fig.\,9  confirm that the normalized initial amplitudes of asymmetries $\rho_s$ and $\rho_n$ indeed reflect the fractions of components with zero and non-zero $B$, respectively. 

It should be noted that the situation is less certain at  small depth due to a decreasing asymmetry of backscattered muons \cite{Morenzoni_2002}, and to lesser extent to reflected muons stopping in the radiation shield. This led to increased error bars for the data obtained at energies $\lesssim$ 5 keV.

In the three panels of Fig.\,9 the volume fractions of the N  and S components in the In-A film are shown vs muon energy $E$. Corresponding average distances $z$ are given at the upper scale of each panel.  The data points depicted as solid circle were extracted from the spectra measured at fixed $H_0$ and varying $E$; open circles are the data obtained from measurements at fixed energies and varying $H_0$. Fig.\,10 shows fractions of regions with $B=0$ outside the film at different distances $z$ from the surface. We note that, as seen from the latter figure,  the areas with zero $B$ outside the sample exist even at the maximum distance (close to 1 $\mu$m) where we probed the field, which is consistent with the above estimate for $L_h$ in Tinkham's scenario. 
  % at low field.  

\begin{figure}
	\includegraphics[width=0.75\linewidth]{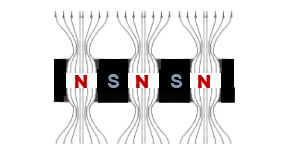}% Here  is how to import EPS art
	\caption{The field distribution and domain shape near the  surface of a sample in the IS at high field. N and S stand for normal and superconducting phases, respectively.}
	\label{fig:epsart}
\end{figure}

Examining the data shown in Fig.\,9, we see that at a low  field ($H_0$ = 44 Oe, the upper panel) the experimental points randomly scatter around nearly  constant values of $\rho_n$ and $\rho_s$ close to 0.5. Taking into account that, as seen from Fig.\,10, the same fraction of regions with zero $B$ is present outside the film at this $H_0$, we conclude that the field lines inside the sample approach the surface being close to parallel and therefore the FDDS at low fields is consistent with the Tinkham's scenario depicted in Fig.\,1B. This explains the successful application of the Tinkham's formula for $L_h$ in aforementioned works \cite{Niedermayer,IS}, where calculations were performed for relatively low fields.  

However at higher field (in the area approximately outlined  by rectangles in Figs.\,7a and 7b) we see that the observed enhancement of $B$ outside the film is accompanied by a decrease of $\rho_n$ near the surface inside it. This leads us to suggest that for high fields the cross sectional domain shape is similar to that shown in Fig.\,1D and therefore FDDS in this range of \textit{perpendicular} fields qualitatively looks as shown in Fig.\,11. 

\begin{figure}
	\includegraphics[width=0.9\linewidth]{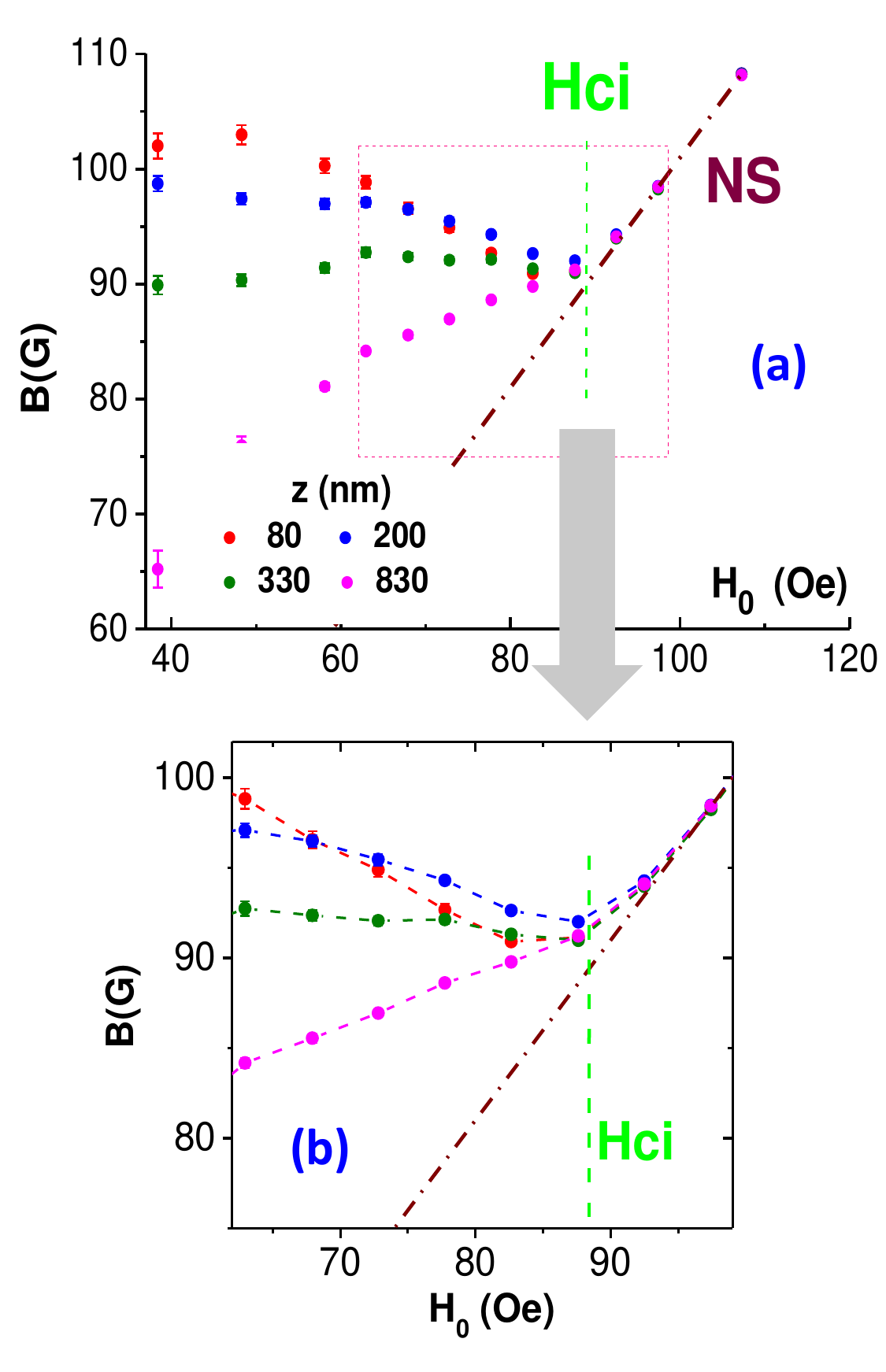}% Here  is how to import EPS art
	\caption{(a) Induction measured outside the In-C film vs  applied field at distances indicated. The data were taken after zero field cooling of the sample. (b) The same data in the upper part of the field range of the IS in an enlarged scale; dashed lines connect the points. The dashed-dotted brown line in both panels is $B(H_0)$ in the normal state (NS). $H_{ci}$ marks the critical field of the IS/NS transition. }
	\label{fig:epsart}
\end{figure}

To verify the results for FDDS obtained with the In-A film, a  similar (but less detailed) experiment was conducted with the In-C film.  Data  for $B$ vs $H_0$ extracted from the spectra measured at fixed distances outside the film are shown in Fig.\,12. The lower panel shows the data at high field in enlarged scale. As one can see, alike for the In-A film, there is also a region in the upper half of the field range of the IS, were the induction $B(z)$ first passes through a maximum before it relaxes down to the applied field far away from the sample. Note that the error bars in the lower panel are barely visible because they are smaller than the size of the symbols. Therefore we conclude that the results obtained for the In-A film reflect a general feature taking place near the surface of superconductors in the IS.

\section{DISCUSSION}

The observation of the near-surface widening of the  S-laminae and corresponding narrowing of their N-counterparts at high applied field testifies that gain in the sample free energy due to the former exceeds losses due to the latter combined with the losses due to the increasing inhomogeneity of the outside field. But then a legitimate question arises: if it is profitable at high fields, why it does not take place at lower field values? 

A possible answer is as follows. As known \cite{Tamm}, the  magnetic field with
induction $B$ in the N domain exerts pressure on the  N/S interface equal to $B^2/8\pi$. A maximum pressure withholding by the N/S interface in type-I superconductors is reached when $B=H_c$ \cite{VK}. It is also known \cite{Egorov,IS}, that in the  IS $B$ in N domains decreases with increasing applied field from $H_c$ down to $H_{ci}$. Therefore, since narrowing of an N-lamina leads to increasing $B$ due to flux conservation, at low $H_0$ (when $B \simeq H_c$) there is no room to make the narrowing profitable. However such an opportunity does appear at high applied field at which $B$ can be significantly (for more than 50\%) less than $H_c$ \cite{IS}. Therefore maximum saving of the condensation energy (proportional to the volume of the S phase) can be reached with a flat interphase boundary (as in Tinkham's scenario in Fig.\,1B) at low field and with widening S-laminae (as in Fig.\,11) at high field.

Another %principal 
question, associated with the FDDS  measured in this work, is related to the theorem of potential  also referred to as Earnshaw's theorem. 

As known,  at steady-state conditions in free space  %(i.e., in absence of currents) 
the Maxwell equations for the magnetic field take the form $\nabla\cdotp \textbf{B}=0$ and $\nabla\times \textbf{B}=0$, which is identical to corresponding Maxwell equations for the static electric field $\textbf{E}$ in free space. % (i.e., in absence of electric charges).   
Hence, at these conditions the magnetic field represents a potential-like field and therefore it  can be described using a magnetic scalar potential $\Psi(\textbf{r})$ ($\textbf{r}$ is a spatial coordinate) for which the Laplace equation $\nabla^2 \Psi=0$ holds (see, \textit{e.g.}, \cite{Jackson}). Then, based on this equation, one can show \cite{Landafshitz_II}  that  $\Psi(\textbf{r})$ can reach an extremal (maximum or minimum) value and therefore $B(=-\bigtriangledown \Psi)$ can be zero only at the edges of a region where there is the field. This is the essence of Earnshaw's theorem  which, thus, stems from solution of the Laplace equation.
	
Under slightly different view-point this theorem can be  considered  coming from the fact that a potential field can be mapped using equipotential lines. Then Earnshaw's theorem states that in free space the equipotential lines of the static magnetic field, like the equipotential lines of the static electric field, can not make closed loops and therefore the field lines can diverge or converge only monotonically not leaving the voids.  

In either way,  \tb{the experimentally obtained field configurations  depicted in  Figs.\,1B (low fields) and 11 (high fields) look} conflicting with  this theorem or, more generally, with solution of the Laplace equation.

\tb{This issue can be resolve as follows. First, we remind that magnetic field is the solenoidal, but not the potential field. Using the mathematical apparatus of potential fields (such as the Laplace equation) for magnetic fields is, in some cases, a convenient procedure, which, however, does not carry a strict physical meaning \cite{Tamm}. Second, the key word in the theorem of potential is ``free space". Normally it is defined as a bound-less space not containing either resting or moving charges. However, in vicinity of material objects this definition is incomplete. Indeed, in vicinity of a normally conducting object (sample) the free space is the space remote from  the sample by a distance greatly exceeding the size of the atoms. At closer distances the Laplace equation is inapplicable due to effects of the irregular fields caused by inhomogeneous microscopic currents beneath the sample surface \cite{Landafshitz_II}. In other words, the applicability of the theorem of potential to the magnetic field is a matter of a boundary condition, which normally does not include an "active" superconductor \cite{Michael}.}

\tb{In case of superconductors in the IS the space effectively bounded by Tinkham's healing length is \textit{controlled} by the superconductor. Indeed, energy of the field in this space is a composite part of the sample free energy \cite{Tinkham,IS,VK}. At this situation specific configuration of the FDDS is dictated by the thermodynamic profitability of the entire sample, which includes the healing space on both sides of the sample. Hence this space is in no means ``free"  and therefore the theorem of potential is inapplicable in this region by definition.} 

As one can see from  the lower panel of Fig.\,6 the studied outside distances from the sample are well within the healing length (beyond this length $B=H_0$). Therefore, the reported field distribution outside the samples is incomparable with the theorem of potential due to inapplicability of this theorem to the range of the studied distances. %, but it sheds  new light at the notion of the free space in vicinity of superconductors. 

\tr{Finally, we note that the considered case of inapplicability of the theorem of potential is not the only one when this theorem fails for the static magnetic fields. For example, according to this theorem a magnetizable body (a system of magnetic dipoles) can not be found in the stable equilibrium in any configuration of the static magnetic fields. However, as it was shown by Braunbeck \cite{Braunbeck} (see also \cite{Brandt}) this is incorrect for diamagnetic bodies. }

\section{SUMMARY AND OUTLOOK}

Eight decades ago Landau for the first time has shown the  determinative role of the near-surface field distribution and of the domain shape for forming the flux structure of the intermediate state in type-I superconductors.  In this work these properties were for the first time measured by low-energy muon spin rotation spectroscopy on pure-limit type-I indium films. It was found that the field-domain configuration proposed  by Tinkham is consistent with our experimental results at low values of the applied field. However at higher fields our observations suggest that the cross-sectional width of the superconducting domains near the sample surface is \textit{widening}, instead of the expected narrowing. Then the field lines emerge from the normal domains converging, and the field outside the sample passes through a maximum before it relaxes to a uniform applied field far away from the sample. There is no reason to believe that similar field/domain configurations are not possible near the surface of type-II superconductors in the mixed state, however details can be different. Verification of these near-surface properties in  type-II superconductors constitutes an interesting problem of fundamental superconductivity which is important for a better understanding of the properties of the mixed state, especially in thin films. 

\begin{figure}
	\includegraphics[width=0.9\linewidth]{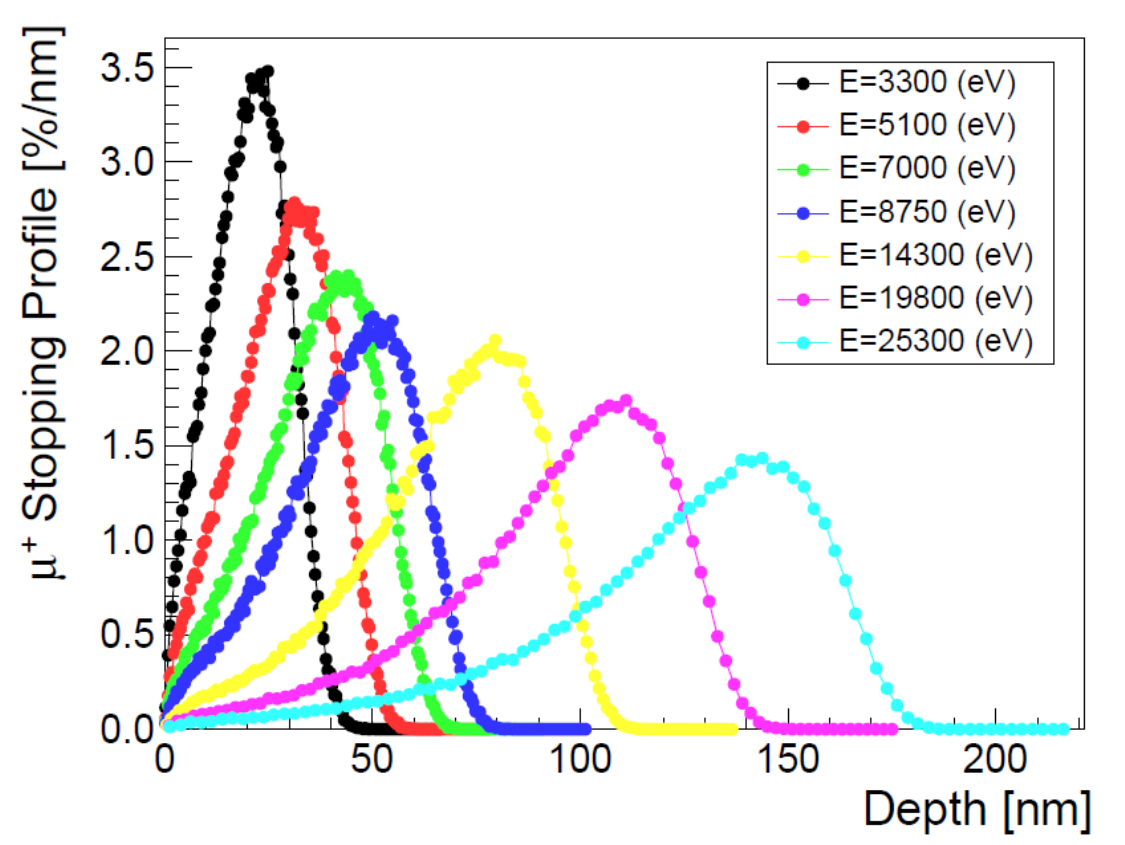}% Here  is how to import EPS art
	\caption{Depth profile of the stopping distances of muons  of different energies in indium. }
	%\label{fig:epsart}
\end{figure}
\begin{figure}
	\includegraphics[width=0.9\linewidth]{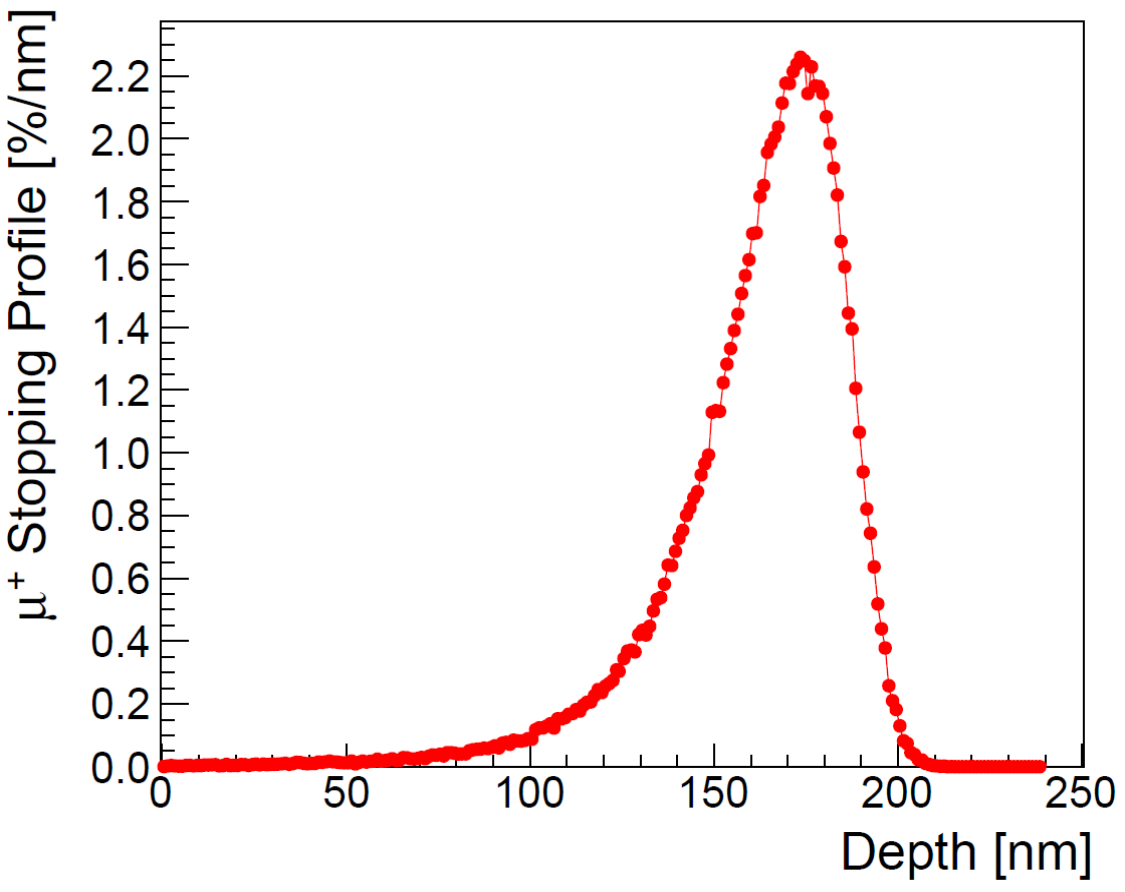}% Here  is how to import EPS art
	\caption{Muon stopping distribution in solid nitrogen for  an implantation energy $E=14.3$keV. }
	%\label{fig:epsart}
\end{figure}
In this work the near-surface properties inside and outside  superconductors with an inhomogeneous flux distribution were measured applying a large scale $\mu$SR facility. As of today, this is the only technique appropriate for such kind of measurements inside the sample. Unfortunately, the main anomalies inside appear very close to the surface, where the accuracy of the $\mu$SR data reduces due to effects mostly associated with the muon backscattering.  
However, on the outer side of the sample, where the field  inhomogeneities are extended over significantly greater length scale,  more detailed and potentially more accurate measurements both with type-I and type-II superconductors can be performed using non-invasive scanning techniques, such as those based on the Hall microprobe, squid-on-tip, or electronic spin resonance of a single nitrogen vacancy center in diamond. We are looking forward to seeing results of such measurement and are ready to share necessary high purity samples. 
\begin{figure}
	\includegraphics[width=0.95\linewidth]{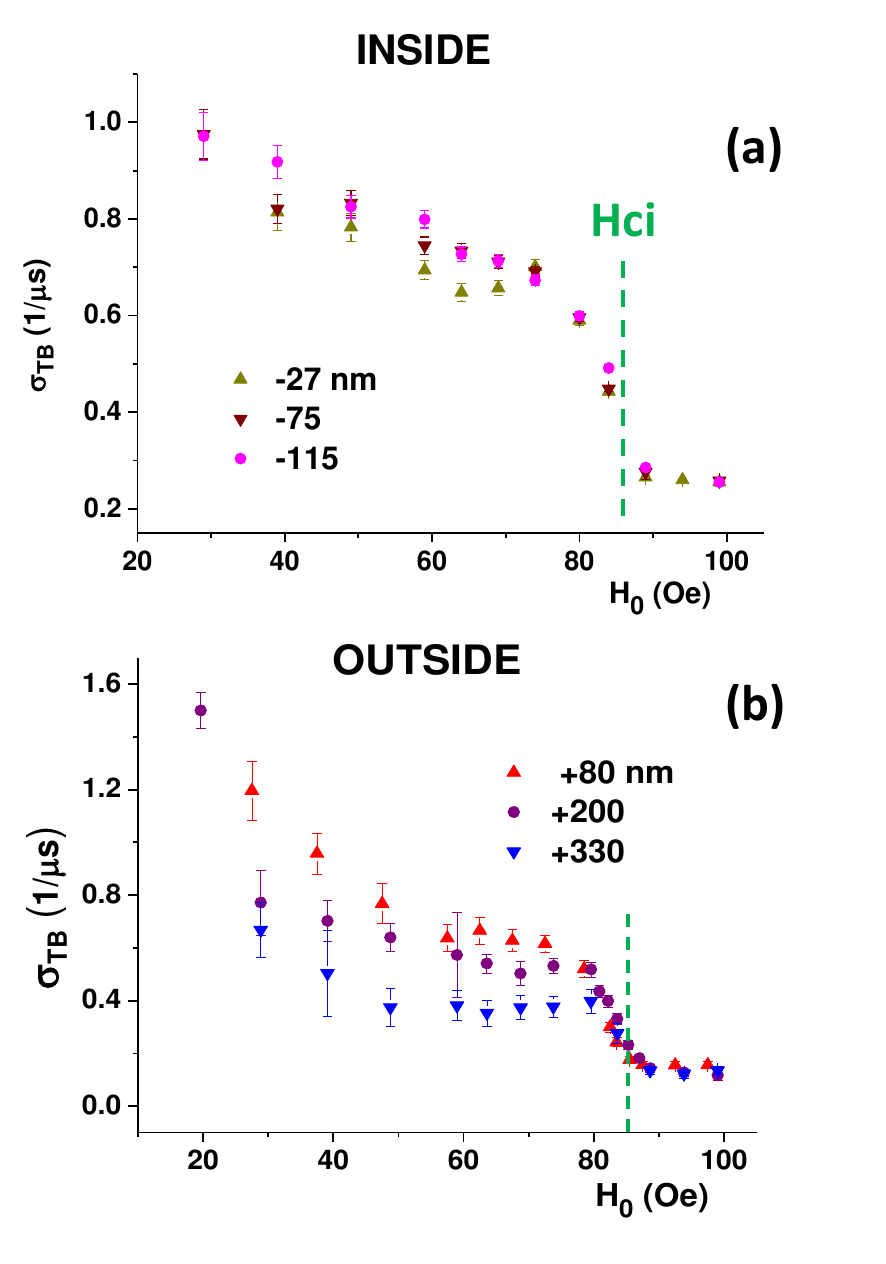}% Here  is how to import EPS art
	\caption{Depolarization rates $\sigma_{\rm TB}$  recorded by the Top and Bottom counters inside (a) and outside (b) of IN-A sample at distances from the surface as indicated. Vertical dashed line marks the critical field $H_{ci}$.}   
	%\label{fig:epsart}
\end{figure}
\begin{figure}
	\includegraphics[width=0.9\linewidth]{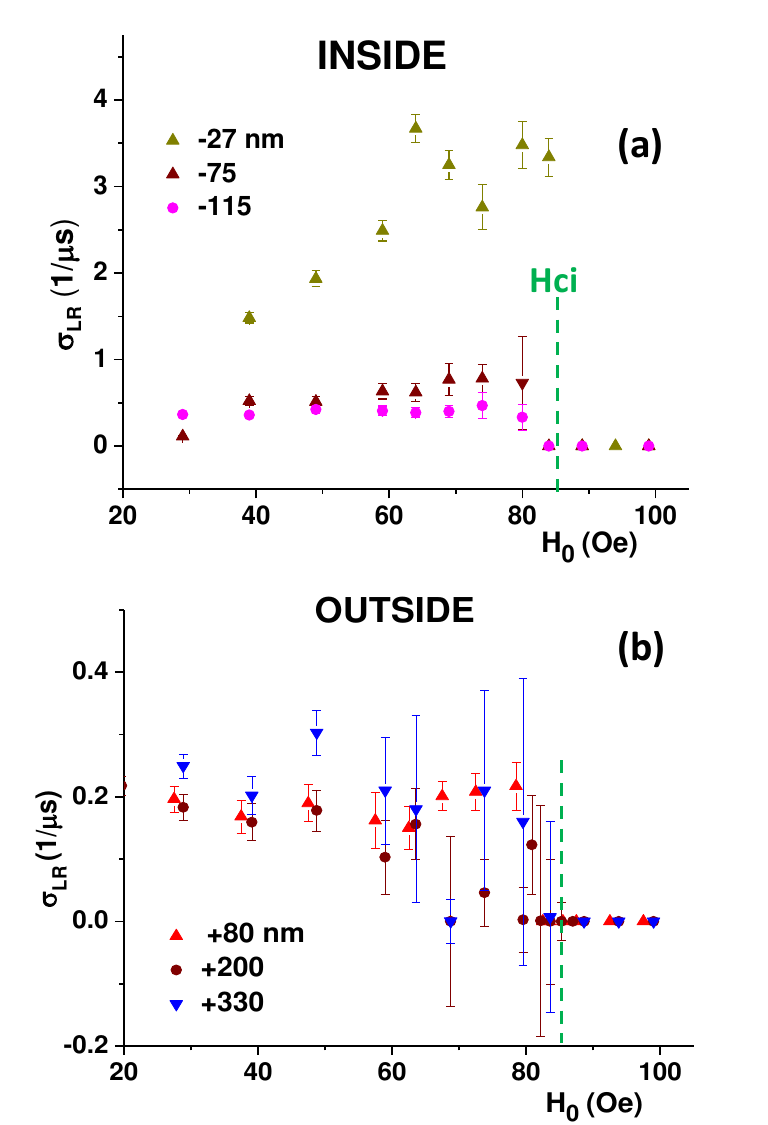}% Here  is how to import EPS art
	\vspace{3mm}\caption{Depolarization rates  $\sigma_{\rm LR}$ recorded by the Left and Right counters inside (a) and outside (b) of  IN-A sample at distances from the surface as indicated. Vertical dashed line marks the critical field $H_{ci}$. }
	%\label{fig:epsart}
\end{figure}

\begin{acknowledgments}
The muon measurements were performed at the Swiss Muon Source  S$\mu$S, Paul Scherrer Institute, Villigen, Switzerland. This work was supported by the National Science Foundation (Grant No. DMR 0904157), by the Research Foundation -- Flanders (FWO, Belgium) and by the Flemish Concerted Research Action (BOF KU Leuven, GOA/14/007) research program. V.K. acknowledges support from the sabbatical fund of the Tulsa Community College. \vspace{3mm}
\end{acknowledgments}

%\appendix*
%\section{Muon Stopping Distribution in Indium and Solid Nitrogen.}

\section*{APPENDIX}
The muon stopping distribution for a given material is  calculated with the Monte Carlo code TRIM.SP  \cite{Eckstein,Morenzoni_2002}. Stopping profiles for various muon implantation energies for indium are shown in Fig.\,13. Fig.\,14 depicts the muon stopping distribution in solid N$_2$ for a muon implantation energy of $E=14.3$ keV.

The average stopping distance (implantation depth)  $\overline{x}(E)$ at given muon energy $E$  is calculated as 

\begin{equation*}
\overline{x}(E)=\int_0^\infty xf(x,E)dx,
\end{equation*}

where $x$ is the stopping distance \label{key}and  $f(x,E)$ is the stopping distances distribution shown in Figs.\,13 and 14.
	
The out-of-plane coordinate $z$ (shown in Fig.\,1B) for  the data obtain with muons implanted into the sample equals $-\overline{x}(E)$; and $z$ for the data measured outside is $z=\Delta - \overline{x}(E)$, where $\Delta$ is the thickness of the nitrogen overlayer. Details about the depth profiles of the low-energy muons are available in \cite{Morenzoni_2002,Morenzoni_2004}.

Figs.\,15 and 16 show representative data for  depolarization rates $\sigma_{\rm TB}$ and $\sigma_{\rm LR}$, respectively, recorded with IN-A sample without (panels (a)) and with (panels (b)) N$_2$ overlayes.

\end{document}